\documentstyle[12pt]{article}
\newcommand{\beq}{\begin{equation}}
\newcommand{\eeq}{\end{equation}}
\newcommand{\beqa}{\begin{eqnarray}}
\newcommand{\eeqa}{\end{eqnarray}}
\newcommand{\ba}{\begin{array}}
\newcommand{\ea}{\end{array}}

\begin{document}

\begin{flushright}
Preprint CAMTP/97-3\\
June 1997\\
\end{flushright}

\vskip 0.5 truecm
\begin{center}
\Large
{\bf  Comment on energy level statistics in the mixed regime}\\
\vspace{0.25in}
\normalsize
Marko Robnik and Toma\v z Prosen\ddag
\footnote{e--mails: robnik@uni-mb.si, prosen@fiz.uni-lj.si}\\
\vspace{0.3in}
Center for Applied Mathematics and Theoretical Physics,\\
University of Maribor, Krekova 2, SLO--2000 Maribor, Slovenia\\
\ddag Department of Physics, Faculty of Mathematics and Physics,\\
University of Ljubljana, Jadranska 19, SLO-1111 Ljubljana, Slovenia\\

\end{center}

\vspace{0.3in}

\normalsize
\noindent
{\bf Abstract.} We comment on the recent paper by Abul-Magd 
({\em J.Phys.A: Math.Gen.} {\bf 29} (1996) 1) concerning the
energy level statistics in the mixed regime, i.e. such having the
mixed classical dynamics where regular and chaotic regions 
coexist in the phase space. We point out that his basic assumption
on the additive property of the level-repulsion function $r(S)$
(conditional probability density) in the sense of dividing 
it linearly into the regular and chaotic part in proportion
to the classical fractional phase space volumes $\rho_1$
and $\rho_2=q$ is {\em not justified}, since among other
things, it relies on the
type of Berry's ergodic assumption, which however is right
only in a homogeneous ensemble of ergodic systems, but not in the
neighbourhood of an integrable system. Thus his resulting
distribution cannot be regarded as a theoretically well
founded object.  We point
out that the semiclassical limiting energy level spacing 
distribution must be of Berry-Robnik (1984) type, and
explain what transitional behaviour of the Brody-type
(with fractional power-law energy level repulsion) we observe
in the near semiclassical regime where effective $\hbar$ is
not yet small enough. Thus we refer to the derivation,
arguments and conclusions in our paper (Prosen and Robnik,
{\em J.Phys.A: Math.Gen.} {\bf 26} (1994) 8059), and explain
again the behaviour in this double transition region.

\vspace{0.6in}

Submitted to {\bf Journal of Physics A: Mathematical and General}
\normalsize
\vspace{0.1in}
  
\newpage


\noindent
In a recent paper (Abul-Magd 1996) offers a new theoretical
energy level spacings distribution for quantal Hamiltonian
systems whose classical dynamics is of the mixed type, i.e.
such having regular regions of invariant tori coexisting in
the phase space (and on the energy surface) with chaotic
regions, a typical KAM scenario. In this comment we want to point
out that his result is not theoretically well founded and is in fact
erronous, and has no other merit than mathematical simplicity,
which however is of course not a sufficient condition for
the scientific merit. Abul-Magd uses the famous Wigner surmise
(Wigner 1956, Brody 1973, Brody et al 1981, Robnik 1984, Bohigas and Giannoni 
1984), which by itself is a sound argument, but
he makes an assumption about the conditional 
probability density $r(S)$ (the so-called level repulsion function),
in conjunction with the Berry-type argument on the ergodicity
of quantal energy spectra in an ensemble of classically
ergodic systems (Berry 1981, 1983, 1985), which is wrong in his context, because it is applied to the
systems that are {\em not} ergodic but close to an integrable system
(KAM type systems). 
\\\\
Quite generally, by knowing $r(S)$ one gets the level spacings 
distribution $P(S)$ at once as $P(S) = r(S) \exp (-\int_{0}^{S}r(x)dx)$.
For example, $r(S)=1$ implies Poisson distribution, $r(S)=\pi S/2$
implies Wigner (2-D GOE), $r(S)\propto S^{\beta}$ implies Brody
distribution and so on. 
\\\\
In order to derive $r(S)$ Abul-Magd refers to the ergodicity
argument by Berry (1981, 1983, 1985) where in calculating the $P(S)$ 
at small $S$ he replaces the average over the energy spectrum by the
average over an ensemble of {\em classically ergodic} systems
parametrized by at least two parameters (because the degeneracies,
the diabolical points, have codimension 2), which is a very
reasonable assumption indeed, and it immediately yields the linear
level repulsion. However, as it has been pointed out by one of
us (Robnik 1984) this ergodicity assumption cannot be applied
in the neighbourhood of an integrable system, simply because there
is no local uniformity in the parameter space, and as we approach
(the coordinates/parameters of) the integrable system we see greater and
greater density of degeneracy points: they are not uniformly
distributed in the space \{parameter A x parameter B x energy\}.
\\\\
Therefore, the ansatz ("the basic assumption") of Abul-Magd
for $r(S)= \rho_1 + \rho_2 S$, where $\rho_1+\rho_2=1$ 
(in his notation $q\equiv \rho_1$), is not justified theoretically,
but is just a guess. In fact it leads to a distribution
function which is mathematically simple, normalized, 
but not its first moment, which is another deficiency of the model. 
\\\\
Further we claim with full theoretical justification that the
correct ultimate semiclassical energy level spacings distribution
is in fact Berry-Robnik (1984). (Similar thinking can be of course
applied to other statistical measures, such as number variance and
delta statistics, etc, see e.g. (Seligman and Verbaarschot 1985)).
This assertion has a sound theoretical foundation. It is based on the picture
in the quantum phase space (the Wigner functions of stationary
eigenstates) in the strict semiclassical limit, $\hbar \rightarrow 0$,
where we observe the condensation of states in volume elements of
order $(2\pi\hbar)^f$, where $f$ is the number of degrees of freedom
(see e.g. Robnik 1988, 1997), on classical invariant objects,
which is the contents of the so-called principle of uniform semiclassical 
condensation.  The prediction agrees with the
rigorous results by Lazutkin (1981, 1991) on splitting the
energy spectra and the eigenstates in regular and irregular levels/states
(qualitatively predicted by Percival (1973)), in the special case 
of convex billiards with smooth boundaries. This has been analyzed also
in (Li and Robnik 1995).
We have at least two special but typical mixed dynamical systems
for which we have demonstrated with a very great accuracy that the
semiclassically limiting statistics is Berry-Robnik, namely in
the quantized compactified standard map (on a torus) (Prosen
and Robnik 1994a,b) and in the 2-D semiseparable oscillator
(Prosen 1995, 1996). In both cases the quantally derived $\rho_1$
agrees with the classical one better than within 1\%. It is
interesting to note that in analyzing the numerical spectra we had
to use the infinitely dimensional GOE statistics on chaotic component
(Wigner distribution = 2-dim GOE was not good enough) in order 
to achieve perfect agreement between the numerical results and
the best fitting Berry-Robnik distribution.
\\\\
Therefore we have full confidence in the correctness
of the asymptotic far semiclassical limit ($\hbar\rightarrow 0$) 
of spectral statistics. 
\\\\
However, before reaching the ultimate semiclassical
limit, in a regime which we call the  near
semiclassical limit, we find phenomenologically significant and
to some extent universal statistical behaviour of energy spectra,
especially in 2-D billiards and elsewhere (Prosen and Robnik 1993,
1994a,b). Namely, we typically observe the fractional power law
level repulsion, $P(S) \propto S^{\beta}$, where the exponent $\beta$
can be anything between  $0$ and $1$ for {\bf OE} statistics, or
$\beta \in [0,2]$ in case of {\bf UE} (broken antiunitary symmetries,
or more generally, complex representations, see (Robnik 1986,
Leyvraz, Schmit and Seligman 1996, Keating and Robbins 1997,
Dobnikar 1996, Robnik and Dobnikar 1997).
It is now qualitatively understood that this statistics is
another manifestation of quantum (dynamical) localization,
i.e. the localization of quantum eigenstates related to
the classical dynamics. In KAM systems we have a theory on $\beta$,
where we derive (Prosen and Robnik 1994b) the scaling law
$\beta = {\rm const.}\hbar$, for sufficiently small $\hbar$.
Furthermore, we have shown that the fractional power-law regime
with the given $\beta$ should be observed for spacings $S$ within the interval
$[\exp(-1/\beta),1]$. Using the above scaling estimates, we see that
the fractional power-law level repulsion is {\em not}  observed 
in the exponentially small interval  $[0,\exp(-{\rm const.}/\hbar)]$.
So, when the {\em effective} Planck constant goes to zero,  
$\hbar \rightarrow 0$, this interval becomes exponentially small
and practically invisible, because there are usually not enough objects there.
The estimate agrees perfectly well with the prediction by Berry
and Robnik (1984) that there is such a exponentially small
region at small $S$ due to the tunneling phenomena. Therefore,
the picture is now fully consistent, and it remains to
explain what behaviour do we predict theoretically in this
exponentially small region. 
\\\\
We know from the elementary thinking that in this region the $P(S)$ must 
behave linearly $P(S) \propto S$, which cannot be predicted semiclassically
(Robnik 1986, Berry 1991, Robnik and Salasnich 1997)
but only quantally (Robnik 1987). The reason is that for very small
spacings the quantum degenerate 2-dim perturbation theory
must be ultimately sufficient, which was demonstrated and argued 
in (Robnik 1987).
Indeed, if one increases the dimensionality of such a model
("Poissson + GOE"), one finds the same linear level repulsion 
law for 3-dim and 4-dim (Izrailev 1993) and for higher dimensions
(Prosen 1993). This quantum mechanical picture explains
the linear level repulsion region, which is exponentially small.
\\\\
The question then is to explain how - in this doubly transitional
regime: mixed dynamics, and transition from near to far semiclassics - 
the Brody-like behaviour goes over into Berry-Robnik behaviour,
as the $\hbar$ tends to $0$.
For this we have no global quantitative theory, except for
the more or less local features described above. Schematically
we show this in figure 1. We also show in figure 2, schematically,
the Brody-like distribution and the Berry-Robnik distribution,
with the indicated (and schematically exaggerated) exponentially
small region of linear level repulsion (the purely quantum regime).
In practice, with actual spectra, it is almost impossible to
detect the exponentially small region, and indeed this has not been
observed until now in any specific system.
\\\\
The existence of the fractional power-law 
level repulsion and Brody-like behaviour is definitely connected
with the existence of (dynamical) localization, which is a topic
of current research (Frahm and Shepelyanski 1997, 
Casati and Prosen 1997, Robnik et al 1997).
Moreover, in ergodic systems, but with very slow diffusion, we also
observe dynamical localization (Prosen and Robnik 1994b, 
which gives rise to the Brody-like 
behaviour, with fractional power-law level repulsion, but here
the picture is much more complicated, and $\beta$ must tend to $1$,
rather than $0$, as $\hbar \rightarrow 0$. 
\\\\
Finally, in regard of phenomenological formulae, we suggest that
the so-called BRB-distribution (Berry-Robnik-Brody, see 
(Prosen and Robnik 1994b)),
which is a two-parameter distribution function, is the best, because
it has the theoretical foundation in the sense that it takes into
account the division of the classical phase space (parameter $\rho_1$),
and the localization of the chaotic states on the (subset of the) chaotic 
regions (whose measure is $\rho_2=1-\rho_1$), 
captured by the level repulsion parameter $\beta$.
Indeed, in our own work (Prosen and Robnik 1994b) we have
confirmed the agreement in billiard systems and in mappings,
and a similar success is reported in the context of theoretical nuclear
spectra by Lopac, Paar and Brant (1996).
\\\\
In conclusion, we propose that there is no place for other limiting
semiclassical energy level statistics than Berry-Robnik (1984),
in systems with mixed classical dynamics (KAM type systems),
while in the transition regime there is evidence and substantial
understanding that outside the exponentially small region of
linear level repulsion due to tunneling, there is the fractional 
power-law level repulsion and Brody-like behaviour with exponent 
$\beta={\rm const.}\hbar$, which goes to zero when $\hbar$ goes to
zero, thereby going over to the Berry-Robnik distribution.
We have explained why the basic assumption of Abul-Magd (1996)
is not justified and therefore any significant agreement of his
results with high-quality spectral data cannot be expected.

\section*{Acknowledgements}
\par
The financial support by the Ministry of Science
and Technology of the Republic of Slovenia is acknowledged with
thanks.

\newpage

\section*{References} 
\parindent=0. pt
Abul-Magd A Y 1996 {\em J.Phys.A: Math.Gen.} {\bf 29} 1
\\\\
Berry M V 1981 {\em Ann. Phys. NY} {\bf 131} 136
\\\\
Berry M V 1983 in {\em Chaotic Behaviour of Deterministic Systems}
(Amsterdam: Norh-Holland) eds G Iooss, R H G Helleman and R Stora,
p 171
\\\\
Berry M V 1985 in {\em Chaotic Behaviour in Dynamical Systems},
ed G Casati (New York: Plenum)
\\\\
Berry M V 1991 in {\em Chaos in Quantum Physics} eds M-J Giannoni,
A Voros and J Zinn-Justin (Amsterdam: North-Holland) 
\\\\
Berry M V and Robnik M 1984 {\em J.Phys.A: Math.Gen.} {\bf 17} 2413
\\\\
Bohigas O and Giannoni M-J 1984 {\em Lecture Notes in Physics} {\bf 209} 
(Berlin: Springer) 1
\\\\
Brody T A 1973 {\em Lett. Nuovo Cimento} {\bf 7} 482
\\\\
Brody T A, Flores J, French J B, Mello P A, Pandey A and Wong S S M 1981
{\em Rev. Mod.  Phys.} {\bf 53} 385
\\\\
Casati G and Prosen T 1997 {\em Phys. Rev. Lett.} in press
\\\\
Dobnikar J 1996 {\em Diploma Thesis}, unpublished, in Slovenian
\\\\
Frahm K and Shepelyanski D 1997 {\em Phys. Rev. Lett.} {\bf 78} 1440
\\\\
Izrailev F M 1993 private communication
\\\\
Keating J P and Robbins J M 1997 {\em J. Phys. A: Math. Gen.} {\bf 30} L177
\\\\
Lazutkin V F 1981 {\em The Convex Billiard and the Eigenfunctions of the 
Laplace Operator} (Leningrad: University Press) in Russian
\\\\
Lazutkin V F 1991 {\em KAM Theory and Semiclassical Approximations
to Eigenfunctions} (Heidelberg: Springer)
\\\\
Leyvraz F, Schmit C and Seligman T H 1996 {\em J. Phys. A: Math. Gen.}
{\bf 29} L575
\\\\
Li Baowen and Robnik M 1995 {\em J.Phys.A: Math.Gen.} {\bf 28} 4483 
\\\\
Lopac V, Paar H and Brant S 1996 {\em Z. Phys. A} {\bf 356} 113
\\\\
Percival I C 1973 {\em J. Phys. B: At. Mol. Phys.} {\bf 6} L229
\\\\
Prosen T 1993 unpublished
\\\\
Prosen T 1995 {\em J. Phys. A: Math. Gen.} {\bf 28} L349
\\\\
Prosen T 1996{\em Physica D} {\bf 91} 244
\\\\
Prosen T 1997a,b {\em Phys.Lett.A} in press
\\\\
Prosen T and Robnik M 1993 {\em J.Phys.A: Math.Gen.} {\bf 26} 2371
\\\\
Prosen T and Robnik M 1994a {\em J.Phys.A: Math.Gen.} {\bf 27} L459
\\\\
Prosen T and Robnik M 1994b {\em J.Phys.A: Math.Gen.} {\bf 27} 8059
\\\\
Robnik M 1984 {\em J. Phys. A: Math.Gen.} {\bf 17} 1049
\\\\
Robnik M 1986 {\em Lecture Notes in Physics} {\bf 263} (Berlin:
Springer) 120
\\\\
Robnik M 1987 {\em J. Phys. A: Math. Gen.} {\bf 20} L495
\\\\
Robnik M 1988 in {\em Atomic Spectra and Collisions in External Fields}
ed K T Taylor, M H nayfeh and C W Clark  (New York: Plenum) pp265-74.
\\\\
Robnik M 1997 {\em Open Systems and Information Dynamics} in press
\\\\
Robnik M and Dobnikar J 1997 to be published
\\\\
Robnik M, Dobnikar J, Liu Junxian and Veble G 1997 work in progress
\\\\
Robnik M and Salasnich 1997 {\em J. Phys. A: Math. Gen.} {\bf 30} 1711
\\\\
Seligman T H and Verbaarschot J J M 1985 {\em J. Phys. A: Math. Gen.} {\bf 
18} 2227
\\\\
Wigner E P 1956 Contribution to {\em "Conference on Neutron Physics
by Time-of-Flight}, Galinburg, Tenessee 1956, reprinted in:
Porter C E Ed. {\em Statistical Theories of Spectra: Fluctuations"}
(Academic Press) 1965
\\\\
\newpage

\section*{Figure captions}

\vspace{0.3in}
\noindent
{\bf Figure 1}: We show the schematic diagramme of the doubly-transition
region: from integrable to ergodic classical dynamics and from
near- (not very small $\hbar$) to far-semiclassics (sufficiently
small $\hbar$).

\vspace{0.3in}
\noindent
{\bf Figure 2}: We show schematically two examples of the Brody-like
level spacings distribution (with higher maximum) and Berry-Robnik type, 
but in both cases indicated the exponentially small (but here
exaggerated)  regime of linear level repulsion (see text).

\end{document}